\documentclass{article}

\usepackage{PRIMEarxiv}
\usepackage[utf8]{inputenc} % allow utf-8 input
\usepackage{caption}
\usepackage{xurl}           % allow long URLs to break across lines
\usepackage{hyperref}       % hyperlinks
\usepackage{booktabs}       % professional-quality tables
\usepackage{amsfonts}       % blackboard math symbols
\usepackage{nicefrac}       % compact symbols for 1/2, etc.
\usepackage{microtype}      % microtypography
\usepackage{lipsum}
\usepackage{fancyhdr}       % header
\usepackage{graphicx}       % graphics
\usepackage{xcolor}         % colors
\usepackage{amsmath}
\graphicspath{{figures/}{media/}{./}}
\usepackage{float}
\usepackage{tabularx}
\usepackage{natbib}
\usepackage{xparse}
\usepackage{multirow}
\usepackage{placeins}
\usepackage{listings}
\lstdefinelanguage{json}{
  basicstyle=\ttfamily\small,
  numbers=left,
  numberstyle=\tiny\color{gray},
  stepnumber=1,
  numbersep=8pt,
  showstringspaces=false,
  breaklines=true,
  frame=single,
  backgroundcolor=\color{gray!5},
  literate=
   *{0}{{{\color{numb}0}}}{1}
    {1}{{{\color{numb}1}}}{1}
    {2}{{{\color{numb}2}}}{1}
    {3}{{{\color{numb}3}}}{1}
    {4}{{{\color{numb}4}}}{1}
    {5}{{{\color{numb}5}}}{1}
    {6}{{{\color{numb}6}}}{1}
    {7}{{{\color{numb}7}}}{1}
    {8}{{{\color{numb}8}}}{1}
    {9}{{{\color{numb}9}}}{1}
    {:}{{{\color{punct}{:}}}}{1}
    {,}{{{\color{punct}{,}}}}{1}
    {\{}{{{\color{delim}{\{}}}}{1}
    {\}}{{{\color{delim}{\}}}}}{1}
    {[}{{{\color{delim}{[}}}}{1}
    {]}{{{\color{delim}{]}}}}{1},
}

\definecolor{numb}{rgb}{0.6,0,0}
\definecolor{punct}{rgb}{0.2,0.2,0.2}
\definecolor{delim}{rgb}{0,0,0.5}

\title{\textbf{\NAMEEPSTEIN}: A Large-Scale Twitter Dataset on the Jeffrey Epstein Case and Its Global Public Discourse (2019–2023)
}

\author{
  Michael Kreil\textsuperscript{1}\\
  \textit{Published as Independent Researcher} \\
  \texttt{paper@michael-kreil.de} \\
  \url{https://michael-kreil.de}
  \And
  Tristan Manfred Stöber\textsuperscript{1}\\
  \textit{Published as Independent Researcher} \\
  \texttt{tristan.stoeber@posteo.net} \\
  \url{https://orcid.org/0000-0003-3853-0608}
  \And
  Daniel Thilo Schroeder\textsuperscript{1}\\
  \textit{Published as Independent Researcher} \\
  \texttt{contact@danielthiloschroeder.org} \\
  \url{https://orcid.org/0000-0003-0125-5243}
}

\newcommand{\NAME}{FUBU}
\newcommand{\NAMEEPSTEIN}{\NAME-EPSTEIN}

\begin{document}
\maketitle
\footnotetext[1]{All authors contributed to this work in their personal capacity and free time. This research is not affiliated with or supported by any of the authors' employers, academic institutions, or privately owned commercial enterprises.}
\renewcommand{\thefootnote}{\arabic{footnote}} % restore normal numbering

\begin{abstract}
The criminal case of Jeffrey Epstein has generated a complex, long-running global discourse on digital platforms, characterized by punctuated attention shocks, conspiracy theories, and blame attribution. To facilitate the computational study of these dynamics, we introduce the \NAMEEPSTEIN\ dataset, a large-scale, multi-dimensional research corpus of 54.38 million Twitter statuses authored by 7.11 million users and collected between August 2019 and April 2023. The source corpus was captured continuously in near-real time and enriched with a directed social contact graph of 37.03 million edge rows and annotations from Qwen2.5-7B-Instruct covering sentiment, conspiracy and misinformation stance, toxicity, moral emotion, and related dimensions. For public distribution, we created a textless, de-identified derivative that retains one row for every deduplicated status, categorical and numeric annotations, coarse temporal information, and 46.15 million internal status relationships. It excludes tweet text, original post and user identifiers, handles, profile fields, exact timestamps, and reverse mappings. The resulting release supports longitudinal content and diffusion analyses while reducing disclosure and platform-content redistribution risks. To request access to raw data for collaborative research under ethical and legal safeguards, contact \href{mailto:fubu.dataset@gmail.com}{\textbf{fubu.dataset@gmail.com}}.
\end{abstract}

\section*{Introduction}
% %%%%%%%%%%%%%%%%%%%%%%%%%%%%%%%%%%%%%%%%%%%%%%%%
The criminal case of Jeffrey Epstein constitutes a highly salient instance of institutional failure and delayed accountability, spanning decades of sexual abuse allegations, elite protection, and contested enforcement outcomes~\cite{Cook2023PedophilesProsecutorsPower}.
Public attention to the case intensified through punctuated \textit{information events} that repeatedly reignited debate about complicity, transparency, and justice. 
These key moments included renewed investigative reporting, major court filings, and a post-mortem focus on facilitators and institutions.
In digital environments, these attention shocks interacted with memetic and conspiratorial framings (e.g., the enduring “Epstein didn’t kill himself” motif), producing a long-running, polarized discourse that blends primary-source fragments (flight logs, dockets, leaked documents) with speculative narratives and networked blame attribution~\cite{SchattoEckrodtCleverFrischlich2024SeedOfDoubt,AttanasioEtAl2025AlgorithmicVisibilityRedditEpstein}.

Recently, the mandated governmental publication of the "Epstein files" ~\cite{DOJ3p5MillionPages2026,DOJEpsteinTransparencyActMemoPDF} has triggered renewed public interest and a resurgence of these investigative and attributional dynamics across digital platforms.

% %%%%%%%%%%%%%%%%%%%%%%%%%%%%%%%%%%%%%%%%%%%%%%%%
\section*{Dataset}
% %%%%%%%%%%%%%%%%%%%%%%%%%%%%%%%%%%%%%%%%%%%%%%%%
To study the formation of this digital discourse, we introduce the \textbf{\NAMEEPSTEIN} dataset, a multi-dimensional archive capturing \textbf{54.38~million Twitter/X statuses} (comprising 4.22M tweets, 42.39M retweets, 6.64M replies, 1.08M quotes, and 51k quote-in-reply posts; see Table~\ref{table:epstein_stats} for exact counts) authored by \textbf{7.11~million unique users} between August 3, 2019 and April 3, 2023. 
The collection was gathered continuously using daily queries based on a fixed set of query terms capturing both core case keywords and prominent conspiratorial or attributional expressions---which could appear as hashtags or as ordinary words (e.g., \texttt{epstein}, \texttt{epsteingate}, \texttt{epsteinblackbook}, \texttt{epsteinsuicidecoverup}, \texttt{clintonbodycount}, \texttt{trumpbodycount}).
The collection is further enriched by an annotated social contact graph of \textbf{37.03~million edges} and LLM-derived content labels across conspiracy, misinformation, and affective dimensions.

By utilizing continuous, near-real-time collection, the private source corpus preserves platform output that was later deleted, edited, or made private.
This approach captures the unfiltered narrative ecology in which claims, memes, and attributional frames initially consolidated.
Furthermore, the integration of longitudinal user profile histories and embedded retweets extending back to 2006 enables fine-grained analyses of how accountability narratives evolve, how evidence fragments circulate and are recontextualized, and how complex social network interactions shape public discourse over time. More broadly, because the collection largely predates the widespread deployment of LLM-based agentic influence systems, it can serve as a historical baseline for comparing human- and bot-mediated misinformation dynamics with emerging forms of synthetic consensus and coordinated AI-swarm manipulation~\cite{SchroederEtAl2026AISwarms}.

To clarify the structural composition of this corpus, we use the term \emph{status} as a generic label for all message types on the platform, which can be public or restricted to followers of the author.
A status is called a \emph{tweet} if it is not directly related to a preceding status. Comments on tweets are called \emph{replies}, which can themselves receive further replies, making it possible to form nested conversation structures often referred to as \emph{threads}.
Both tweets and replies are shareable.
If a status is shared without additional commentary, we refer to it as a \emph{retweet}; if it is shared with an attached comment, we refer to it as a \emph{quote tweet}.
Since a retweet reproduces the content of the original status, it can be interpreted as an act of amplification (and, in many contexts, agreement).
Because replies can also be retweeted, a reply may appear both as part of the original conversation and as the root of a new subgraph.

Because multiple keyword queries were issued in parallel during collection, the same status could be returned more than once.
As a consequence, the raw dataset \emph{contains duplicates} that must be filtered out in downstream processing.
However, because collection occurred in near-real-time and ran without interruption, this redundancy inherently captures statuses before they were removed and allows us to track \emph{the evolution of individual tweets} over time.
For example, by comparing different snapshots of a post as it appeared at different collection points, it is possible to observe changes in mutable metadata or edited content.

Likewise, each status object contains the author’s profile data as returned at collection time (e.g., screen name, display name, profile description, and other metadata).
This implies that, across the full dataset, it is possible to reconstruct \emph{the evolution of user profiles over time} by comparing successive snapshots for the same user.

Finally, while the collection window begins at the stated start date, tweet timestamps in the corpus may predate the first day of collection when a retrieved reply, quote, or retweet references an older tweet; in such cases, the referenced tweet is captured as part of the nested payload.

% %%%%%%%%%%%%%%%%%%%%%%%%%%%%%%%%%%%%%%%%%%%%%%%%
% % DATASET AND LABEL STATISTICS
% %%%%%%%%%%%%%%%%%%%%%%%%%%%%%%%%%%%%%%%%%%%%%%%%
\begin{table}[H]
\centering
\small
\setlength{\tabcolsep}{5pt}
\renewcommand{\arraystretch}{1.04}
\caption{Overview of the private research corpus. Counts were generated from the deduplicated V3 data and statistics on March 31, 2026.}
\label{table:epstein_stats}
\begin{tabularx}{\linewidth}{@{}XrX@{}}
\toprule
\textbf{Measure} & \textbf{Exact value} & \textbf{Scope} \\
\midrule
Collection period & Aug.\ 3, 2019--Apr.\ 3, 2023 & Daily collection window \\
Status timestamp range & Mar.\ 21, 2006--Apr.\ 3, 2023 & Includes embedded referenced statuses \\
Total statuses & 54,375,364 & Deduplicated records \\
Tweets & 4,217,545 & 7.8\% of statuses \\
Retweets & 42,394,124 & 78.0\% of statuses \\
Replies & 6,636,902 & 12.2\% of statuses \\
Quotes & 1,075,740 & 2.0\% of statuses \\
Quotes in reply & 51,053 & 0.1\% of statuses \\
Unique observed authors & 7,107,409 & Authors of retained statuses \\
Unique source texts & 13,983,232 & Units submitted for labeling \\
Contact-graph edge rows & 37,028,919 & User-level derived graph \\
Top month & 9,573,314 & August 2019 statuses \\
Top language & 49,597,846 & English (91.2\%) \\
\bottomrule
\end{tabularx}
\end{table}

The documented search terms, added on August 13, 2019 and unchanged thereafter, were
\texttt{clintonbodycount}, \texttt{clintoncrimefamily}, \texttt{clintonsbodycount},
\texttt{epstein}, \texttt{epsteinblackbook}, \texttt{epsteincoverup},
\texttt{epsteingate}, \texttt{epsteinmurder}, \texttt{epsteinsuicide},
\texttt{epsteinsuicidecoverup}, \texttt{epsteinsuicided}, \texttt{epsteinunsealed},
\texttt{epstien}, \texttt{jefferyepstein}, \texttt{jeffreyepstein}, and
\texttt{trumpbodycount}.

% %%%%%%%%%%%%%%%%%%%%%%%%%%%%%%%%%%%%%%%%%%%%%%%%
\subsection*{Data and Code Availability}\label{sec:data_code_availability}
% %%%%%%%%%%%%%%%%%%%%%%%%%%%%%%%%%%%%%%%%%%%%%%%%
The anonymized V4 linked release is publicly available on Zenodo at \url{https://doi.org/10.5281/zenodo.21621944}. The release does not redistribute raw Twitter/X content.

A public Python reader for the V3 binary format and the anonymized Parquet release is available at \url{https://github.com/DTSchroeder/large-twitter-data-processor}. Collection code for querying the Twitter API, managing keyword lists, and validating file integrity is available at \url{https://github.com/MichaelKreil/twitter-analysis}.

To request access to raw data for collaborative research under ethical and legal safeguards, contact \href{mailto:fubu.dataset@gmail.com}{\textbf{fubu.dataset@gmail.com}}.
% %%%%%%%%%%%%%%%%%%%%%%%%
\subsection*{Content and Label Statistics}
\begin{figure}[!p]
  \centering
  \includegraphics[width=\linewidth]{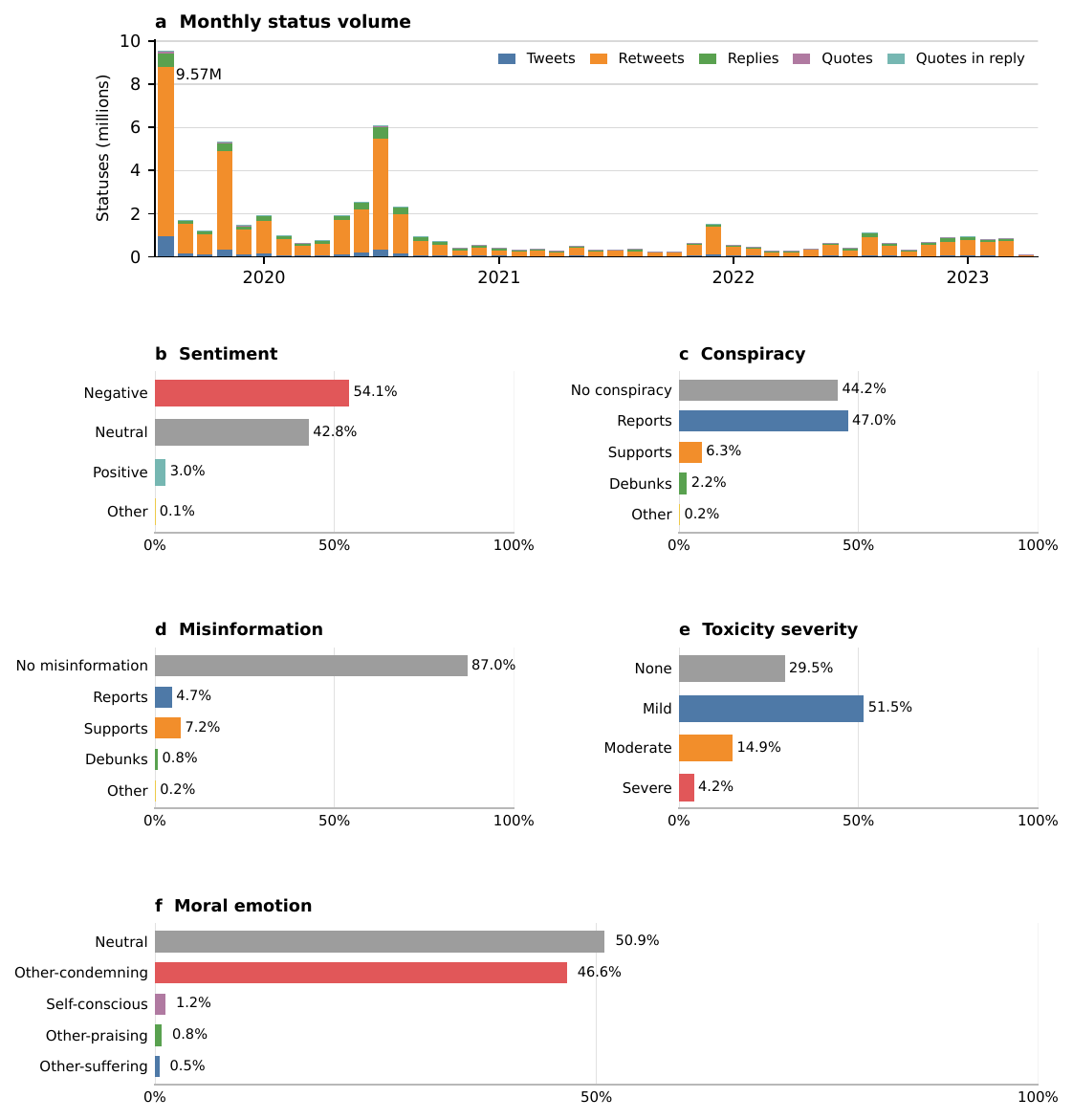}
  \caption{Corpus activity and selected Qwen2.5-7B-Instruct label distributions. Panel a shows monthly statuses by type during the primary collection window; panels b--f show distributions across 13,981,306 valid labels assigned to unique source texts. Off-schema raw variants are grouped as ``Other.'' The 1,926 parsing errors (0.014\% of all annotation rows) are excluded.}
  \label{fig:activity_labels}
\end{figure}

% %%%%%%%%%%%%%%%%%%%%%%%%
Figure~\ref{fig:activity_labels} summarizes monthly status volume and five central annotation dimensions. The corpus peaks sharply in August 2019 and is dominated by retweets throughout the collection period. Across 13,981,306 valid Qwen labels, sentiment is primarily negative (54.1\%) or neutral (42.8\%). Conspiracy-related claims are reported in 47.0\% of labels, supported in 6.3\%, and debunked in 2.2\%. Misinformation support appears in 7.2\% of labels. Toxicity is most often mild (51.5\%), while moral-emotion labels are concentrated in neutral (50.9\%) and other-condemning (46.6\%) categories. The additional political and mobilization dimensions are sparser: ideology is detected in 8.1\% of valid labels, partisanship in 9.6\%, and mobilization in 2.4\%.

% %%%%%%%%%%%%%%%%%%%%%%%%
\subsection*{Public Release Format}
% %%%%%%%%%%%%%%%%%%%%%%%%
The public release is approximately 1.34\,GB and uses Apache Parquet for the three record-level tables summarized in Table~\ref{table:public_release}. It contains one row for every deduplicated status, including retweets, replies, quotes, and quotes in reply. Researchers can join \texttt{statuses.annotation\_id} to \texttt{annotations.annotation\_id}, and can join both endpoint columns in \texttt{relations.parquet} to \texttt{statuses.record\_id}. All identifiers are release-specific and carry no platform meaning.

\begin{table}[H]
\centering
\small
\setlength{\tabcolsep}{4pt}
\renewcommand{\arraystretch}{1.05}
\caption{Files in the anonymized V4 linked release candidate.}
\label{table:public_release}
\begin{tabularx}{\linewidth}{@{}p{0.27\linewidth}rX@{}}
\toprule
\textbf{File} & \textbf{Rows} & \textbf{Contents} \\
\midrule
\texttt{statuses.parquet} & 54,375,364 & Release record and user IDs, total chronological order, coarse dates, status type, annotation ID, and language. \\
\texttt{annotations.parquet} & 13,983,232 & Categorical and numeric Qwen labels plus six independent moral-emotion scores; no source text or free-text model fields. \\
\texttt{relations.parquet} & 46,150,021 & Internal retweet, reply, and quote links between release record IDs. \\
\texttt{aggregates/} & --- & Temporal, language, label, diffusion, and network summaries. \\
\texttt{manifest.json} & --- & Provenance, schemas, counts, exclusions, and anonymization decisions. \\
\texttt{validation\_report.json} & --- & Schema, identifier, temporal, and referential-integrity checks. \\
\texttt{checksums.sha256} & --- & SHA-256 checksums for every release file. \\
\bottomrule
\end{tabularx}
\end{table}

% %%%%%%%%%%%%%%%%%%%%%%%%%%%%%%%%%%%%%%%%%%%%%%%%
\section*{Methods}
% %%%%%%%%%%%%%%%%%%%%%%%%%%%%%%%%%%%%%%%%%%%%%%%%
\subsection*{Data Collection}
\begin{figure}[H]
  \centering
  \includegraphics[width=0.42\linewidth]{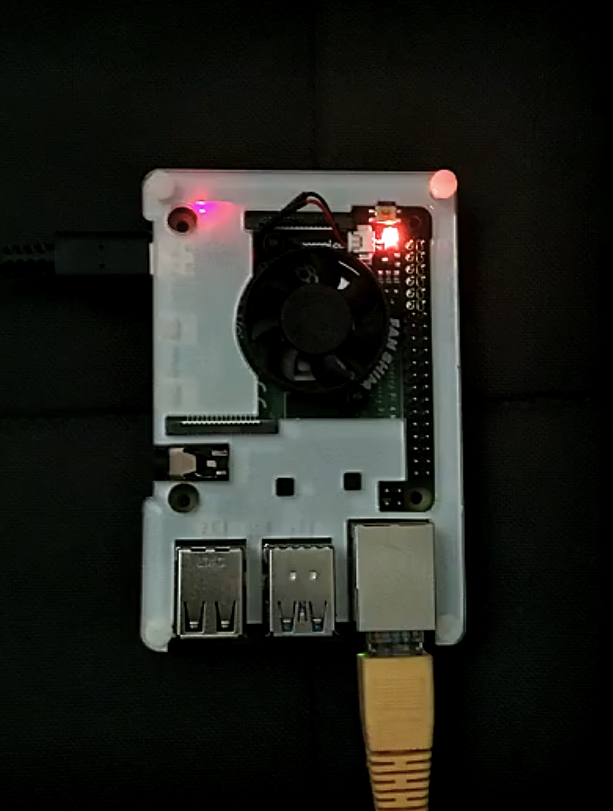}
  \caption{Scraping setup for Twitter data collection.}
  \label{fig:setup}
\end{figure}

Data were collected continuously (24/7) via the official Twitter Search API v1.1 from \textbf{August 3, 2019} to \textbf{April 3, 2023}. 
We utilized a scalable social media mining framework similar to~\cite{burchard2020scalable} to manage available API tokens and rate limits, deploying the collection scripts on a Raspberry~Pi~4 (see Figure~\ref{fig:setup}). 
The daily queries were driven by the fixed terms listed in the Dataset section, which were \emph{not adjusted} after the documented update on August 13, 2019.

During collection, we stored the raw JSON objects exactly as returned by the API without intermediate filtering. To maintain chronological integrity and facilitate downstream reproducibility, the incoming data stream was batched and saved as \emph{one compressed archive file per day}.

% %%%%%%%%%%%%%%%%%%%%%%%%
\subsection*{Preprocessing Pipeline}\label{sec:preprocessing}
% %%%%%%%%%%%%%%%%%%%%%%%%
We developed a custom Java-based preprocessing pipeline to transform the raw JSON stream into compact binary representations suitable for large-scale analysis. The pipeline operates in five stages and is designed to process datasets that exceed main memory capacity on a single node (up to 1.9\,TB heap on 2\,TB RAM nodes). Processing is performed using RxJava-based reactive streams with up to 200 parallel workers.

\paragraph{Stage~1: Field Extraction.}
The pipeline performs seven sequential passes over all raw \texttt{.jsonl.xz} files, one for each entity type: URLs, hashtags, symbols, tweet texts, languages, user locations, and user mentions. During each pass, the pipeline decompresses the XZ archives, parses each JSON line, and recursively ``unfolds'' nested status objects (i.e., \texttt{retweeted\_status}, \texttt{quoted\_status}) into independent records. For each entity type, the pipeline collects all unique string values and assigns each a compact integer identifier, producing a bidirectional string$\leftrightarrow$ID mapping. These mappings are serialized as memory-mapped binary files, split at 2\,GB boundaries for portability.

\paragraph{Stage~2: Compact Status Creation.}
Using the field-to-ID mappings from Stage~1, the pipeline re-reads all raw files, parses each JSON object using an internal BSON parser, and constructs typed compact status records. The parser classifies each unfolded status into one of five canonical types: \texttt{Tweet}, \texttt{Retweet}, \texttt{Reply}, \texttt{Quote}, or \texttt{QuoteInReply}---based on the presence of \texttt{retweeted\_status}, \texttt{quoted\_status}, and \texttt{in\_reply\_to\_status\_id} fields. Each compact record stores: the status ID, user ID, integer-encoded arrays of hashtags/URLs/symbols/mentions, location ID, text ID, language ID, timestamp, engagement counts (retweet and like counts), and a sensitivity flag. All string-valued fields are replaced by their integer identifiers, reducing storage from $\sim$294\,GB (uncompressed JSON) to a compact binary representation.

\paragraph{Stage~3: Deduplication.}
Because the collection system issued multiple keyword queries in parallel, the raw corpus contains duplicate records (i.e., the same status returned by different queries). The pipeline extracts all unique status IDs into a hash set, then filters the compact status stream to retain only the first occurrence of each ID, eliminating duplicates.

\paragraph{Stage~4: Chronological Sorting.}
After deduplication, all compact statuses are sorted by their \texttt{created\_at} timestamp, producing a single chronologically ordered binary archive. This enables efficient temporal slicing and time-series analyses.

\paragraph{Stage~5: Social Contact Network Construction.}
The pipeline constructs a directed social contact graph from the deduplicated, sorted statuses. Edges are created based on direct user interactions: retweets, replies, quotes, quote-in-reply links, and \texttt{@}-mentions in replies and quotes. The private graph is exported as \texttt{edge\_list\_labeled.csv}, where each row records source and target user IDs, contact and mention counts, and associated label vectors. Table~\ref{table:network_stats} reports aggregate network statistics for \NAMEEPSTEIN{}; the public release contains the corresponding non-individual summaries in \texttt{aggregates/network\_*.csv}.

\begin{table}[H]
\centering
\small
\setlength{\tabcolsep}{4pt}
\renewcommand{\arraystretch}{1.03}
\caption{Aggregate network statistics for \NAMEEPSTEIN{}. Cleaned graph metrics exclude negative-ID rows and self-loops.}
\label{table:network_stats}
\begin{tabularx}{\linewidth}{@{}p{0.28\linewidth}rX@{}}
\toprule
\textbf{Metric} & \textbf{Value} & \textbf{Definition} \\
\midrule
Raw directed edge rows & $37{,}028{,}919$ & Rows in \texttt{edge\_list\_labeled.csv}. \\
Excluded rows & $276{,}551$ & $35{,}433$ negative-ID rows and $241{,}118$ self-loops. \\
Cleaned directed edges & $36{,}752{,}368$ & Edges retained for graph-level summaries. \\
Edge-list endpoints & $7{,}139{,}833$ & Unique source/target user IDs after cleaning. \\
Largest weak component & $6{,}909{,}712$ & $96.8$\% of cleaned edge-list endpoints. \\
Mean out-degree & $5.15$ & Cleaned directed graph. \\
Directed density & $7.21\times10^{-7}$ & Excluding self-loops. \\
Directed reciprocity & $1.83$\% & $335{,}973$ reciprocal unordered dyads. \\
\bottomrule
\end{tabularx}
\end{table}

\subsection*{Release Anonymization and Validation}
\label{sec:anonymization}
The V4 public package is generated from the deduplicated, chronologically sorted source data through a separate allowlist-based export. It is designed to avoid redistribution of Twitter/X content and to reduce re-identification risk in accordance with the content-redistribution and privacy principles of the X Developer Policy~\cite{XDeveloperPolicy2026}. The export applies the following transformations:

\begin{enumerate}
    \item It assigns independent random permutations of contiguous, zero-based identifiers to statuses, users, and annotations. The randomization seed is not retained, and no reverse mapping is written.
    \item It excludes original status, user, and interaction-target IDs; tweet text and internal text IDs; handles, mentions, hashtags, URLs, symbols, locations, profile fields, source-client strings, free-text model outputs, exact timestamps, and target-user identifiers.
    \item It retains a total \texttt{chronological\_index}, month-level dates for all rows, and day-level dates only for statuses within the primary collection window. This preserves before/after order without publishing exact posting times.
    \item It retains only status relationships whose source and target both occur in the release. This yields 42,394,124 retweet, 2,671,109 reply, and 1,084,788 quote links. A further 4,016,846 replies and 42,005 quotes to external targets are omitted rather than assigned identifiers.
\end{enumerate}

An automated validator enforces the published schemas, verifies that release identifiers are unique and contiguous, confirms that every retained relationship joins to valid status rows and matches the source status type, checks temporal coverage, rejects original-ID columns and symbolic links, and produces file checksums. This design removes direct platform identifiers and text, but it does not make re-identification impossible: large or publicly documented cascades may still have recognizable structural and temporal fingerprints. The release therefore remains subject to institutional disclosure review before deposit.

\subsection*{Text Labeling}
\label{sec:labeling}

To enrich the dataset with content-level annotations, we performed automated labeling of all \textbf{13,983,232 unique tweet texts} (after deduplication) using a large language model. We deployed \textbf{Qwen2.5-7B-Instruct}~\cite{qwen2.5} via eight parallel \textbf{vLLM}~\cite{kwon2023vllm} inference servers, each running on a dedicated NVIDIA H200 GPU. Labeling was performed as a single self-contained Slurm job: the job script launches eight vLLM server processes as background tasks within the same allocation, waits for all servers to pass health checks, and then runs an asynchronous Python labeling client that distributes requests across the eight servers via round-robin scheduling with a concurrency of 48 outstanding requests per server.

\paragraph{Labeling Schema.}
Each unique text is submitted to the model with a structured prompt requesting a JSON response containing topic, stance, toxicity, affective, political, mobilization, and source-likelihood dimensions:

\begin{enumerate}
    \item \textbf{Sentiment}: \texttt{positive}, \texttt{negative}, or \texttt{neutral}.
    \item \textbf{Conspiracy}: A structured object with \texttt{type} (\texttt{supports\_conspiracy} for texts that actively endorse a conspiracy theory, \texttt{reports\_conspiracy} for texts that neutrally discuss conspiracy theories, \texttt{debunks\_conspiracy} for explicit corrections, or \texttt{no\_conspiracy}), a free-text \texttt{topic}, and a \texttt{confidence} score ($0$--$1$).
    \item \textbf{Misinformation}: Analogous structure with \texttt{type} (\texttt{supports\_misinfo} for texts that spread false information as if true, \texttt{reports\_misinfo} for texts that discuss misinformation without endorsing it, \texttt{debunks\_misinfo} for explicit corrections, or \texttt{no\_misinfo}), \texttt{topic}, and \texttt{confidence}.
    \item \textbf{Toxicity}: A \texttt{severity} level (\texttt{none}, \texttt{mild}, \texttt{moderate}, \texttt{severe}) and category scores for hate speech, harassment, threat, and profanity.
    \item \textbf{Moral emotion}: A primary emotion label such as \texttt{neutral}, \texttt{other\_condemning}, \texttt{self\_conscious}, \texttt{other\_praising}, or \texttt{other\_suffering}, plus confidence.
    \item \textbf{Ideology and partisanship}: Detection flags, ideological orientation, targeted partisan group where present, and confidence scores.
    \item \textbf{Mobilization}: Whether the text contains a call to action and, if so, the mobilization type.
    \item \textbf{Persuasive framing}: Persuasive strength and broad narrative frame (e.g., justice or corruption).
    \item \textbf{Bot/spam likelihood}: A coarse likelihood label distinguishing likely human, uncertain, and likely bot/spam text.
    \item \textbf{Primary topic}: A free-text label identifying the main topic of the text.
    \item \textbf{Notable features}: A short free-text tag for the most salient stylistic or rhetorical feature, such as sarcasm, breaking news, or a call to action.
\end{enumerate}

\paragraph{Exact Labeling Prompt.}
The fixed system prompt used for every text is reproduced verbatim below. The
corresponding tweet text, truncated to its first 2,000 characters, was supplied
separately as the user message.

\begin{lstlisting}[
  basicstyle=\ttfamily\scriptsize,
  breaklines=true,
  columns=fullflexible,
  frame=single,
  xleftmargin=0.5em,
  xrightmargin=0.5em
]
You are a text labeling assistant. Given a social media text, respond ONLY with a JSON object (no other text) containing these fields:

{
  "sentiment": "positive" | "negative" | "neutral",
  "conspiracy": {
    "type": "supports_conspiracy" | "reports_conspiracy" | "debunks_conspiracy" | "no_conspiracy",
    "topic": "<brief topic if applicable, else null>",
    "confidence": 0.0-1.0
  },
  "misinformation": {
    "type": "supports_misinfo" | "reports_misinfo" | "debunks_misinfo" | "no_misinfo",
    "topic": "<brief topic if applicable, else null>",
    "confidence": 0.0-1.0
  },
  "toxicity": {
    "severity": "none" | "mild" | "moderate" | "severe",
    "hate_speech": 0.0-1.0,
    "harassment": 0.0-1.0,
    "threat": 0.0-1.0,
    "profanity": 0.0-1.0
  },
  "bot_likelihood": "likely_bot" | "likely_human" | "uncertain",
  "primary_topic": "<main topic of the text>",
  "ideology": {
    "detected": true | false,
    "orientation": "left" | "right" | "anti-establishment" | "unclear" | null,
    "confidence": 0.0-1.0
  },
  "partisanship": {
    "detected": true | false,
    "target": "democrat" | "republican" | "both_parties" | "other_political_group" | "unclear",
    "confidence": 0.0-1.0
  },
  "persuasive_intent": {
    "strength": "strong" | "weak" | "none"
  },
  "frame": {
    "primary": "public_safety" | "rights_and_freedom" | "economic" | "public_health" | "election_integrity" | "corruption" | "identity" | "nationalism" | "security" | "justice" | "elite_vs_people" | "media_critique" | "geopolitics" | "other" | "unclear"
  },
  "mobilization": {
    "present": true | false,
    "type": "protest" | "vote" | "boycott" | "donate" | "share_information" | "join_movement" | "punitive_action" | "other" | "unclear"
  },
  "moral_emotion": {
    "primary": "other_condemning" | "other_praising" | "other_suffering" | "self_conscious" | "neutral",
    "confidence": 0.0-1.0
  },
  "notable_features": "<most notable characteristic in max two words, e.g. 'sarcasm', 'breaking news', 'emotional appeal', 'call to action', 'personal story'>"
}

Key distinctions:
"supports_conspiracy" = actively promotes/endorses a conspiracy theory
"reports_conspiracy" = neutrally reports on or discusses a conspiracy theory without endorsing
"debunks_conspiracy" = explicitly rejects or corrects a conspiracy theory
"supports_misinfo" = actively spreads false or misleading information as if true
"reports_misinfo" = neutrally reports on or discusses misinformation without endorsing
"debunks_misinfo" = explicitly rejects or corrects misinformation

Moral emotion categories (based on moral foundation theory):
"other_condemning" = expresses moral outrage, blame, or disgust toward others' actions/character
"other_praising" = praises or admires others' moral behavior or character
"other_suffering" = expresses empathy, compassion, or concern for others' pain or hardship
"self_conscious" = expresses guilt, shame, or embarrassment about one's own actions
"neutral" = no strong moral emotion detected

Persuasive intent:
"strong" = actively tries to change beliefs or behavior using rhetorical devices, emotional appeals, calls to action, or argumentative framing
"weak" = expresses an opinion or stance but without deliberate persuasive techniques
"none" = purely informational, descriptive, or conversational with no intent to influence

JSON only, no explanation.
\end{lstlisting}

\paragraph{Independent Moral-Emotion Scores.}
Each unique source text was also scored with the six-class \texttt{Chaeyoon/ELECTRA-Moral-Emotion-ENG} classifier\footnote{\url{https://huggingface.co/Chaeyoon/ELECTRA-Moral-Emotion-ENG}}. The public annotation table retains separate probabilities for other-condemning, other-praising, other-suffering, self-conscious, non-moral emotion, and neutral, together with the highest-scoring class and its confidence. These scores are not treated as interchangeable with the Qwen labels: among the 12,096,885 annotation rows referenced by at least one released status, raw highest-class agreement is 47.7\%, increasing to 59.3\% when the classifier's non-moral-emotion class is mapped to Qwen's neutral class.

The model's raw JSON response is parsed into flattened V3 fields, and each labeled text is stored privately as a single JSON line in \texttt{labels\_full.jsonl}. Texts for which the model returns malformed JSON or encounters an error are recorded with an \texttt{error} field; for the \NAMEEPSTEIN{} dataset, only 1,926 out of 13,983,232 labels (0.014\%) resulted in errors. The labeling process supports automatic resumption from partial output, enabling fault-tolerant execution across multi-day runs. The public V4 release contains only the allowlisted categorical and numeric annotation fields described above; source text and free-text model outputs are excluded.

% %%%%%%%%%%%%%%%%%%%%%%%%%%%%%%%%%%%%%%%%%%%%%%%%
\section*{Related Datasets}
% %%%%%%%%%%%%%%%%%%%%%%%%%%%%%%%%%%%%%%%%%%%%%%%%
The Jeffrey Epstein case has generated an unusually rich ``data exhaust'' across social media discourse, investigative/journalistic document archives, and legal and administrative releases. This makes the Epstein case a useful testbed for studying how high-profile criminal cases evolve online, how accountability narratives form, and how conspiratorial framings (e.g., “Epstein didn’t kill himself”) propagate~\cite{SchattoEckrodtCleverFrischlich2024SeedOfDoubt}.

A major body of related resources consists of primary documents released through official channels. The U.S. Department of Justice maintains an \emph{Epstein} portal with disclosure pages and multiple “Data Set” collections released under transparency requirements, providing a structured corpus of case-connected materials that can be used to study how document releases punctuate attention cycles and reshape claims-making online~\cite{DOJEpsteinLibrary,DOJEpsteinDataSet12Files}. In addition, DOJ has published redacted \emph{Maxwell Interview} transcripts and associated files, offering a time-stamped textual record that can be linked to shifts in public interpretation and narrative framing~\cite{DOJMaxwellInterviewLanding}.

Complementing these official sources, widely circulated “network artifacts” such as flight log documents are available as official PDFs (e.g., exhibits associated with U.S.~v.~Maxwell) and are also mirrored in public document repositories such as DocumentCloud~\cite{DOJFlightLogReleasedUSvMaxwellPDF,DocumentCloudEpsteinPrivatePlaneFlightLogs}. These semi-structured artifacts are central to online “association” claims and name-centric secondary investigations, making them useful reference corpora for analyzing how evidence fragments are selected, recontextualized, and mobilized in social media discourse.

Community repackagings transform raw case materials into analysis-ready formats. Kaggle hosts datasets branded as ``Epstein files'' alongside derived tables and notebooks, illustrating how legal and journalistic artifacts are normalized into ML/NLP-friendly corpora~\cite{KaggleEpsteinFilesJazivxt}. Together, these external document corpora provide primary-source anchors that complement \NAMEEPSTEIN. By contextualizing social media discourse against discrete document drops, researchers can study how evidence fragments circulate, evolve, and sustain accountability or conspiratorial narratives in digital spaces.

Beyond document archives, specific social media datasets have been constructed to examine the digital fallout of the Epstein case across platform environments. 
Schatto-Eckrodt et al.\ compiled an 8-million-post cross-platform corpus (Twitter, Reddit, Gab, 4chan) to analyze the immediate genesis of conspiracy narratives following Epstein's death~\cite{SchattoEckrodtCleverFrischlich2024SeedOfDoubt}. 
More recently, Attanasio et al.\ examined platform-level visibility shocks by gathering 2.28~million posts from Reddit's \texttt{r/conspiracy} to evaluate user retention, toxicity, and semantic integration among newcomers~\cite{AttanasioEtAl2025AlgorithmicVisibilityRedditEpstein}.
While these studies provide valuable insights into acute platform shocks and multi-platform dynamics during 2019, their datasets are restricted in scope or unavailable for download due to platform sharing terms. 
In contrast, \NAMEEPSTEIN\ contributes a publicly distributable, textless multi-year derivative equipped with LLM-derived annotations and internal status relationships, enabling researchers to study the long-term evolution and persistence of public discourse without receiving the underlying post text or platform identifiers.
\FloatBarrier

%%%%%%%%%%%%%%%%%%%%%%%%%%%%%%%%%%%%%%%%%%%%%%%%
% REFERENCES
%%%%%%%%%%%%%%%%%%%%%%%%%%%%%%%%%%%%%%%%%%%%%%%%
\bibliographystyle{unsrt}  
\bibliography{references}  

%%%%%%%%%%%%%%%%%%%%%%%%%%%%%%%%%%%%%%%%%%%%%%%%
% END DOCUMENT
%%%%%%%%%%%%%%%%%%%%%%%%%%%%%%%%%%%%%%%%%%%%%%%%
\end{document}